\documentstyle[11pt]{article}

\begin{document}

\baselineskip = 18pt plus 1pt minus 1pt

\title{Flavour Dependence of Form Factors in Heavy Meson Decays}

\author{FE Close\\
\makebox[5cm]{}
\\
{\em Rutherford Appleton Laboratory,}\\
{\em Chilton Didcot, Oxon OX11 0QX, Great Britain.}
\\
\\
 A Wambach\\
\makebox[5cm]{}
\\
{\em Theoretical Physics,}\\
{\em Department of Physics,} \\
{\em 1 Keble Road, OXFORD, OX1 3NP, Great Britain.} \\
\\
\\
RAL--94--041 \\
\\
OUTP -- 94 09P
\\
\\
}
\vspace{2cm}
\date{ April 1994}

\maketitle

\begin{abstract}
\noindent Recently we have shown that due account of Wigner spin rotations is
needed to match the ISGW model   consistently onto HQET. We now discuss the
flavour dependence of this procedure. We find that for finite mass ``heavy''
quark the universal function $\xi(y=1) \approx 1$ but the slope is proportional
to the energy of the spectator quarks, and ratios of form factors are corrected
in a way that QCD sum rules seem to require.
Predictions for heavy quarks, such as in $B_c \rightarrow \psi(\eta_c)$ and to
mixed systems, such as $D\rightarrow K$ and $B \rightarrow \pi$ show a
systematic mass dependence that may be confronted with data.
\end{abstract}
\pagestyle{empty}

\newpage
\pagenumbering{arabic}
\pagestyle{plain}
The ISGW model \cite{isgur} is widely used in describing heavy quark
transition form factors at low (zero) recoil ($y=v\cdot v' \approx 1$).
In recent works \cite{li}--\cite{wambach2} we have shown  that
matching this model (and indeed any quark model) onto Heavy Quark
Effective Theory (HQET)\cite{neubert} requires considerable care even when $M_Q
\rightarrow \infty$,  due to the nontrivial recoil and spin rotation structure
for light spectator antiquark systems. When these Wigner rotations are
consistently accounted for we found that the model with parameters determined
from a fit to heavy--light spectroscopy, describes dynamical transitions not
just at zero recoil but also is consistent with the leading behaviour for non
zero recoil.

In the present note we extend this analysis,  motivated by the
decay $B_c \rightarrow J/\Psi (\eta_c)e \nu$ where both active and spectator
quarks are relatively heavy.  Although $M_b \rightarrow \infty$ is still
assumed, consistency requires that mass corrections for the $c$--quark are
included, both in its role as active participant and, for the $\bar{c}$, as
spectator.
 The explicit derivation of these corrections
then provides us with a tool for calculating transition elements for
$B \rightarrow D$, $D \rightarrow K$ and even $B \rightarrow \pi$ in a limited
kinematic range.

Our analysis shows that in these cases where
HQET fixes the $y=1$ value of a particular form factor to be non-zero, the
corrections for finite mass active quarks are
small, as is the dependence on the mass of the spectator
antiquark. However, for cases where form factors vanish in the $M_Q \rightarrow
\infty$ zero recoil limit these corrections can
be large. Furthermore the slopes show a strong (approximately linear)
dependence on the mass of the spectator quark.

We begin by extending ref.\cite{close} to the case of large, but
finite, mass active quark, in particular including the Wigner rotation
of its spin. With the formalism set up in previous works
\cite{close,wambach1} this is achieved immediately. For the general case we
consider a meson where the mass of the (active) quark and (spectator)
antiquark are different, in this case denoted by $m_1$ and $m_2$.

In the rest frame the meson wavefunction is written in terms of the quark
and antiquark spinors in the form $u(-\vec{k})\bar{v}(\vec{k})$, where
$\vec{k}$ is the relative momentum, and the energy
of the first (second) quark is $\omega_{1(2)} = \sqrt{m_{1(2)}^2+\vec{k}^2}$.
This then yields:
\begin{equation}
M(v=(1, \vec{0})) = (m_1+\not{\!K})\left( \begin{array}{lr} 0&X\\
                             0&0 \end{array} \right)
(m_2-\not{\!k}) [4m_1(m_1+\omega_1)m_2(m_2+\omega_2)]^{-\frac{1}{2}}
\end{equation}
where $K_\mu$ denotes the four momentum of quark 1, and $X$ is either
$-1$ for the pseudoscalar  or $\vec{\sigma} \cdot \vec{\epsilon} $
 for the vector meson. The equation can be simplified if the internal
momenta $K$ and $k$ are decomposed into the direction of $v$ and
orthogonal to it, i.e. $K_\mu = K_l*v_\mu + K_{t\mu}$ where
$K_l=K\cdot v$ (and analogously for $k$).

Next the wavefunction is boosted to an arbitrary
velocity $v$, where, since $K_t=-k_t$, we obtain for the pseudoscalar meson:
\begin{equation}
M(v)  = (m_1+K_l-\not{\!k_t}) \frac{1+\not{\!v}}{2} \gamma_5
(m_2+k_l - \not{\!k_t}) [4m_2m_1(m_1+K_l)(m_2+k_l)]^{-\frac{1}{2}}
\end{equation}
(for the vector meson,  $\not{\!\epsilon}$ replaces $\gamma_5$).

In the limit where the spectator (anti)quark does not change its
momentum, the matrix element for the
pseudoscalar to pseudoscalar transition takes the following form:
\begin{equation}
\begin{array}{l}
\langle P(v') | V_\mu | P(v) \rangle  =  \xi_1(y) (v+v')_\mu +
\xi_2(y) (v-v')_\mu \\
\\
 =  \int{\rm d}^3k_t{\rm d}^3k_t' \delta(k_t-k_t') {\rm Tr}
[(m_2+k'_l-\not{\!k'_t})\gamma_5\frac{(1+\not{v'})}{2}(m_1+K'_l -
\not{\!k'_t}) \Gamma_{\mu}
\frac{(1+\not{v})}{2}\gamma_5]\phi^{*}(k_t')
\phi(k_t)\\
\end{array}
\end{equation}
where $\phi(k_t)$ is the momentum distribution whose  explicit form was
discussed in \cite{close,wambach1}; as in those references  we  employ
gaussian wavefunctions
with coupling strengths calculated in \cite{isgur,lusignoli}. The
exponential term is then multiplied
by an additional factor $\sqrt{\frac{k_l}{k_0}}$ in order to render
the calculation frame--independent.
The normalization factor (inverse square root factor in eq.(2)) is subsumed
within the wavefunction term $\phi^*$ (and $\phi$ respectively).

The evaluation of $\xi_1$ and $\xi_2$ is done  in the rest frame of
the meson after the decay. The agreement with the result in the rest
frame before the decay was explicitly checked. For this calculation
it is convenient to use:
\begin{equation}
k'_t = (k'_t\cdot v) \frac{v-yv'}{1-y^2} + \kappa
\end{equation}
with $y=v \cdot v'$, and the integration over $\kappa$ is zero.
For example the result for $\xi_1$ is to order $(y-1)$:
\begin{equation}
\begin{array}{ccl}
\xi_1(y) & = & \int{\rm d}^3k' (\beta_1\beta_2\pi)^{-\frac{3}{2}}
e^{(-k^{'2}A)}
\sqrt{\frac{m_1+\omega_1}{2m_1}}  \\
\\
& & \times (1 + 2(y-1)[\frac{1}{4}- \frac{\omega_2}{4(m_2+\omega_2)} +
\frac{k^{'2}}{24(m_2+\omega_2)^2} - \frac{k^{'2}}{24\omega_2^2} -
\frac{3m^2+5k^{'2}}{6\beta_1^2} +
\frac{m_2^2k^{'2}+k^{'4}}{6\beta_1^4}])
\end{array}
\end{equation}
where $A^{-1}=2(\beta_1^2+\beta_2^2)$.
A similar although more laborious calculation provides the form
factors for the pseudoscalar to vector meson transition:
\begin{equation}
\begin{array}{ccc}
\langle V(v',\epsilon) | A_\mu | P(v) \rangle & = & \rho_1(y)
\epsilon^*_\mu + \rho_2(y) (\epsilon^* \cdot v) v_\mu + \rho_3(y)
(\epsilon^* \cdot v) v'_\mu \\
\\
\langle V(v',\epsilon) | V_\mu | P(v) \rangle & = & i \eta(y)
\epsilon_{\mu \alpha \beta \gamma} v^\alpha \epsilon^{*\beta} v^{'
\gamma}\\
\end{array}
\end{equation}

HQET determines the value of these form factors in the $m_1
\rightarrow \infty$ limit, namely:
\newline $\xi_1 = \rho_1/(y+1) = -\rho_3 =
\eta = \xi$ and $\xi_2 = \rho_2 = 0$, where $\xi$ is the Isgur-Wise
function which is normalized to 1 at $y=1$, the zero--recoil point.

Luke's theorem \cite{luke} states that at this zero--recoil point
$\xi_1$ and $\rho_1$ receive corrections only
to order $1/M^2$. From the structure of the expression in (5) it is
apparent that the corrections in the mass of the active quark ($m_1$)
are indeed to order $1/m_1^2$ for $\xi_1(y=1)$. The same result
applies for $\rho_1$. The other form factors are all corrected  by
$1/m_1$  terms.

The origin of the  absence of $1/M$ corrections for only two
form factors  in the quark model description of meson
transitions is the following. The current employed during the
transition has the form $\bar{u} \Gamma_\mu u$; $\Gamma_\mu$ is
diagonal  for $\gamma_0$ and $\vec{\gamma} \gamma_5$ and it is only in
these two cases that the upper and lower components of $u$
and $\bar{u}$ do not get mixed, consequently $1/M$ terms do not
appear. In the zero--recoil frame the expressions for these currents
are:
\begin{equation}
\begin{array}{ccc}
\langle P(v) | \Gamma_0 | P(v) \rangle & = & 2v_0 \xi_1(y=1)\\
\\
\langle V(v,\epsilon) | \vec{A} | P(v) \rangle & = & \rho_1(y=1)
\vec{\epsilon}^*\\
\end{array}
\end{equation}
Therefore $\xi_1$ and $\rho_1$ are protected for corrections in
$1/m_1$.

Let us now consider mass effects for the nonperturbative corrections. In the
ISGW model
of present interest, these are subsumed in a phenomenologically
inspired  non--relativistic Coulomb plus linear potential
\cite{isgur} and   variational solutions to the Schr{\"o}dinger problem
were then found based on harmonic oscillator wavefunctions. The
parameter in this gaussian wavefunction which was  optimized by a fit to
spectroscopy  is the oscillator strength \linebreak
 $\beta \equiv \sqrt{m \Delta E}$  for particle mass $m$ with excitation energy
$\Delta E$. Empirically   $\Delta E$ is approximately flavour independent
\cite{pdb}. In Table 1 we display  the oscillator strengths for various flavour
systems, as calculated in
\cite{isgur,lusignoli}, which confirms the approximate flavour independence of
$\Delta E$ when the reduced mass $m=m_1m_2/(m_1+m_2)$ is used.

\begin{figure}[h]
\begin{center}
\begin{tabular}{|c||c|c|c|c|c|c|c|}  \hline
meson & $B_c$ & $B_s$ & $B$ & $J/\Psi$ & $D$ & $K$  & $\pi$ \\ \hline
$\beta$ (GeV) & 0.82  & 0.51  & 0.41 & 0.66 & 0.39 & 0.34 & 0.31  \\ \hline
$\Delta E_1$  (GeV)  & 0.40  & 0.47  & 0.51 & 0.26 & 0.46 & 0.35 & 0.29  \\
\hline
$\Delta E_2$  (GeV)  & 0.53  & 0.52  & 0.54 & 0.51 & 0.55 & 0.56 & 0.58  \\
\hline
\end{tabular}
\end{center}
{\bf Table 1: }
{\it In row 2 different coupling strengths for different mesons
as given in \cite{isgur,lusignoli} are displayed. In row 3
$\Delta E_1=\beta^2/m$ is determined where $m$ is the mass of the light
constituent (anti)quark. $\Delta E_2$ is calculated with $m$ being the reduced
mass of the
system. Quark masses employed are the constituent masses: $m_u$=0.33GeV,
$m_s$=0.55GeV, $m_c$=1.7GeV and $m_b$=5.12GeV. Note the stability of
$\Delta E_2$.}
\end{figure}

In the $M_Q \rightarrow \infty$ limit, the reduced mass $m$ becomes that of the
light (anti) \linebreak quark. However,
for a consistent description of $1/M$ effects, the
dependence of \linebreak  $\beta$ on the heavy quark mass has to be taken into
account.
In particular, with  \linebreak $\beta=\sqrt{m_1m_2\Delta E/(m_1+m_2)} \approx
\sqrt{m_2\Delta E}(1-m_2/(2m_1))$, effects in $1/M$ are entering the scene.
However, in the zero recoil limit the form factors $\xi_1$ and
$\rho_1$ are still protected against  picking up
nonperturbative $1/M$ corrections, due to the
normalization of the Isgur--Wise function. In the general case, where
the wavefunction for meson 1(2) gets corrected by $1/M_{1(2)}$ terms,
the overall correction to a form factor has the structure:
\begin{equation}
\Delta_\xi = \int {\rm d}^3 k[ \Phi^*(k,\beta_2)
\frac{d\Phi(k,\beta_1)}{d\beta_1}|_{\beta_1=\beta_2}\Delta(\beta_1) +
\frac{d\Phi^*(k,\beta_2)}{d\beta_2}|_{\beta_1=\beta_2}\Phi(k, \beta_1)
\Delta(\beta_2)]
\end{equation}
However, this form factor is normalized at zero recoil
\cite{close1,isgur1} in the infinite mass limit where
$\beta_1=\beta_2=\beta$:
\begin{equation}
\int {\rm d}^3 k \Phi^*(k,\beta)\Phi(k,\beta) = 1
\end{equation}
This result is independent of $\beta$ and therefore
\begin{equation}
\int {\rm d}^3 k \Phi^*(k,\beta)
\frac{d\Phi(k,\beta)}{d\beta} = 0\\
\end{equation}
so that $\Delta_\xi=0$ in eq.(8). A similar conclusion has been made
in ref.\cite{amundson}.

For the following numerical results we performed the calculation in
the limit where the active heavy quark {\it before} the decay is
infinitely heavy (in general the bottom quark), however after the
decay it picks up mass corrections (usually the charm quark). In table
2 the results for a particular case, namely $B \rightarrow D^{(*)}$ are
displayed (i) in the HQET limit and (ii) with mass corrections for the
charm quark in the $D$ meson. To see the dependence of the numerical
values on the mass of the spectator quark we display in row 4 the
same form factors for the decay  $B_s \rightarrow D_s^{(*)}$.

\begin{figure}[h]
\begin{center}
\hspace{-1cm}
\begin{tabular}{|c||c|c|c|c|c|c|}  \hline
form factor & $\xi_1$ & $-\xi_2$ & $\rho_1$ & $\rho_2$ & $-\rho_3$ &
$\eta$  \\ \hline
HQET-limit & 1.00 - 1.40$\epsilon$ & 0 & 2.00 - 1.81$\epsilon$ & 0 &
1.00 - 1.40$\epsilon$ & 1.00 - 1.40$\epsilon$ \\ \hline
finite $m_c$ & 1.01 - 1.43$\epsilon$ & 0.10 - 0.12$\epsilon$ & 2.01 -
1.69$\epsilon$ & 0.17 - 0.32$\epsilon$ & 0.98 - 1.30$\epsilon$ & 1.16
- 1.62$\epsilon$ \\ \hline
$B_s \rightarrow D_s^{(*)}$ & 1.00 - 1.64$\epsilon$ & 0.14 -
0.18$\epsilon$ & 2.00 - 2.08$\epsilon$  & 0.22 - 0.42$\epsilon$ & 0.98
- 1.48$\epsilon$ & 1.19 - 1.91$\epsilon$ \\ \hline
\end{tabular}
\end{center}
{\bf Table 2: }{\it In this table the numerical results for the
different form factors for $B\rightarrow D^{(*)}$, defined in eq.(6),
to order (y-1) are displayed ($\epsilon \equiv (y-1)$). Row 2
shows the results in the HQET limit.
Row 3 includes $1/m_c$ corrections and in row 4 the spectator
antiquark is the strange antiquark instead of the up or down.}
\end{figure}

In general the data behave in the following way: If HQET predicts a
nonzero value for a form factor at a special point, i.e. the
zero--recoil point, the modifications induced by mass effects are not
significant. Furthermore, a variation of the spectator quark mass
changes this result only slightly. However, values which are  not
 restricted by HQET (such as the slope of the form factor, coefficient of
$\epsilon$) do show a very strong dependence on the
spectator quark mass; for example, the slope of $\xi_1$ turns out to be nearly
proportional to the mass
of the spectator quark (see figure 1). This finds its cause in the
non--relativistic
prediction, where $\rho^2 = \frac{m^2}{2\beta^2} = \frac{m}{2\Delta E}$,
together with the empirical flavour independence of $\Delta E$. Similar results
have been noted by Voloshin \cite{voloshin} in a different context.
Deviations  to this linear dependence on the spectator quark mass are
induced by relativistic effects.

In figure 2 we plot the form factor $\xi_1$ as a function of $(y-1)$ for
different meson transitions in a limited range  where $\xi_1$ exhibits  a
linear dependence in $(y-1)$. Note that the slopes for $B \rightarrow \pi$, $D
\rightarrow K$ and $B \rightarrow D$ are approximately parallel, reflecting
their common light spectator (anti)quark;  $D_s \rightarrow \bar{s}s$ is
analogously parallel to $B_s \rightarrow D_s$ though with a larger slope due to
the heavier strange quark mass and $B_c \rightarrow \eta_c$ has the steepest
slope due to the massive charmed spectator.

Comparison with experiments are shown in figure 3.

Effects induced by finite mass corrections are usually discussed in
the context of  the functions $R_1$ and $R_2$ which were  proposed by
Neubert \cite{neubert3}  as a measure of symmetry breaking effects.
They read (in our parametrization of the form factors):
\begin{equation}
\begin{array}{rcl}
R_1(y) & = & 2\frac{\eta(y)}{\rho_1(y)}(1+\frac{1}{2}(y-1))\\
\\
R_2(y) & = & -2\frac{\rho_2(y) + r\rho_3(y)}{\rho_1(y)}(1+\frac{1}{2}(y-1))\\
\end{array}
\end{equation}
where $r = M_{D^*}/M_B$. In the HQET limit $R_1=R_2=1$, however finite
mass corrections as well as QCD correction terms modify this result.

In the  ISGW model, the ratios  $\eta(y)/\rho_1(y)$ and $(\rho_2(y) +
r\rho_3(y))/\rho_1(y)$ are constant, so that the $y$ dependences of $R_1$ and
$R_2$ are that of the common  factor  $(1+\frac{1}{2}(y-1))$. Explicit
evaluation of the ratios leads to:
\begin{equation}
\begin{array}{rcl}
R_1(y) & = & 1.01 + 0.50(y-1) + {\rm O}((y-1)^2)\\
\\
R_2(y) & = & 0.91 + 0.45(y-1) + {\rm O}((y-1)^2)\\
\end{array}
\end{equation}
In our computation the velocity dependence is more subtle. We find
\begin{equation}
\begin{array}{rcl}
R_1(y) & = & 1.15 - 0.07(y-1) + {\rm O}((y-1)^2)\\
\\
R_2(y) & = & 0.91 + 0.04(y-1) + {\rm O}((y-1)^2)\\
\end{array}
\end{equation}
which in sign as well as in magnitude approaches the estimate based on QCD sum
rules, given in \cite{neubert}:
\begin{equation}
\begin{array}{rcl}
R_1(y) & = & 1.35 - 0.22(y-1) + {\rm O}((y-1)^2)\\
\\
R_2(y) & = & 0.79 + 0.15(y-1) + {\rm O}((y-1)^2)\\
\end{array}
\end{equation}

The remaining small discrepancy can be partially explained by noting that $R_1$
receives substantial short--distance corrections proportional to
$\alpha_s(m_c)$ which are not included in the quark model.

We have shown so far the application of the consistent model for space
like form factors, where it appears be a  success. There is
much concern about the flavour (mass) dependence of timelike form factors,
namely the decay constant of
heavy mesons. HQET predicts that the decay
constant has to scale like $M^{-1/2}$. In particular this implies:
$f_B \sqrt{M_B} = f_D \sqrt{M_D}$.

However, lattice calculations \cite{lattice} determine $f_B = 180 \pm
40 $MeV and $f_D = 200 \pm 30 $MeV, which differs from the HQET
prediction:
$f_D = \sqrt{\frac{M_B}{M_D}} f_B = 300 \pm 70$ MeV. QCD sum rules
 imply a similar behaviour \cite{dominguez}.

In the  quark model formalism of this paper, the decay constant is to be
calculated via:
\begin{equation}
\begin{array}{llc}
\langle 0 | A_\mu | P(v) \rangle & \propto & \int
{\rm Tr}[(m_1+\not{\!K_t})\frac{(1+\not{v})}{2}\gamma_5
(m_2-\not{\!k}) \gamma_{\mu}\gamma_5]
[4m_2m_1(m_1+K_l)(m_2+k_l)]^{-\frac{1}{2}}
\phi(k_t)\\
\\
& \propto & v_\mu \int [(m_1+\omega_1)(m_2+\omega_2) + k_t^2]
[4m_2m_1(m_1+K_l)(m_2+k_l)]^{-\frac{1}{2}} \phi(k_t)\\
\\
& \propto & v_\mu \int [ \sqrt{1+\frac{\omega_2}{m_2}} -
\frac{1}{m_1}(\sqrt{\frac{\omega_2}{m_2}-1}*\frac{|k|}{2})]\phi(k_t) +
0(\frac{1}{m_1^2})
\end{array}
\end{equation}
The Wigner--rotation of the spin of the heavy quark thus does correct
the HQET expression, decreasing the value of the decay
constant and thereby bringing $f_B$  and $f_D$ closer to each other.
The numerical importance of the $1/m_1$ correction term is in this
model 3\% for the B-meson and 15\% for the D-meson. This effect alone,
however, is too small to explain the large deviation from the HQET
limit as calculated in lattices and QCD sum rules (see above).
So a description of the (timelike) decay systematics remains elusive in this
approach, whereas the spacelike shows agreement with available data.
Data on $B_c \rightarrow \psi(\eta_c)$ at one extreme and  $B\rightarrow \pi$
at the other will show the range of applicability of this model in the
spacelike domain. If it turns out to be successful, then the problems in the
timelike region will be highlighted.

This feature in the timelike region may be due to the specific probing of the
wavefunction at one point ($r\rightarrow 0$), in contrast to spacelike form
factors which ``average'' over the whole wavefunctions.
 But in any case it is intersting to
note that the corrections induced by the Wigner rotation of the spin
of the heavy quark do correct the decay constant in the right
direction, which suggests that relativistic effects may dominate the decay
amplitude and therefore lie beyond the spectroscopic--based non relativistic
models.

\newpage

%%%%%%%%%%%%%%%%%%%%%%%%%%%%%%%%%%%%%%%%%%%%%%%%%%%%%%%%%%%%%%%%%%%%%%%%%
\begin{figure}[t]
\input epsf
%\newpage
\vspace*{-14cm}
\epsfbox{rho.ps}
\vspace*{-0cm}
 {\it{ {\rm [1]} The figure shows the dependence of the charge radius $\rho^2$
on the mass of the spectator antiquark ($m$). (i) in the HQET-limit and (ii)
with a finite charm quark mass, ($m_c=1.7$ GeV).}}
\end{figure}
%%%%%%%%%%%%%%%%%%%%%%%%%%%%%%%%%%%%%%%%%%%%%%%%%%%%%%%%%%%%%%%%%%%%%%%%%
%%%%%%%%%%%%%%%%%%%%%%%%%%%%%%%%%%%%%%%%%%%%%%%%%%%%%%%%%%%%%%%%%%%%%%%%%
\begin{figure}[h]
\input epsf
\newpage
\vspace*{-15cm}
\epsfbox{xi1.ps}
\vspace*{0cm}
 {\it{ {\rm [2]} The form factor $\xi_1$ is plotted for different
meson transitions. From top to bottom they are: $B \rightarrow \pi$
(short dash), $D \rightarrow K$ (dot -- short dash), $D_s \rightarrow
\bar{s}s$ (long dash), $B_s \rightarrow D_s$ (dotted), $B \rightarrow
D$ (solid), $B_c \rightarrow \eta_c$ (dot -- long dash)}}
\vspace{1cm}
\end{figure}
%%%%%%%%%%%%%%%%%%%%%%%%%%%%%%%%%%%%%%%%%%%%%%%%%%%%%%%%%%%%%%%%%%%%%%%%%
%%%%%%%%%%%%%%%%%%%%%%%%%%%%%%%%%%%%%%%%%%%%%%%%%%%%%%%%%%%%%%%%%%%%%%%%%
\begin{figure}[t]
\input epsf
%\newpage
\vspace*{-14cm}
\epsfbox{argm.ps}
\vspace*{1.5cm}
 {\it{ {\rm [3]} Comparism of the model with experimental data from ARGUS
\cite{argus} for $|V_{bc}|$ in the infinite mass limit  and  with a finite
charm quark mass. Note that the mass dependence is nugatory.}}
\end{figure}
%%%%%%%%%%%%%%%%%%%%%%%%%%%%%%%%%%%%%%%%%%%%%%%%%%%%%%%%%%%%%%%%%%%%%%%%%

\end{document}